\documentclass[preprint]{ptephy_v1}

\preprintnumber{IU-TH-18} 
\usepackage{hyperref}

\usepackage{amsmath}
\usepackage{amssymb}

\usepackage{graphicx}
\usepackage{color}

\usepackage{float}



\begin{document}

\title{Reality from maximizing overlap in the periodic complex action theory}


\author{Keiichi Nagao${}^{1,2,*}$} 

\author{Holger Bech Nielsen${}^{2,*}$}

\affil{${}^{1}$Faculty of Education, Ibaraki University, Bunkyo 2-1-1 , Mito, 
310-8512 Japan 
}

\affil{${}^{2}$Niels Bohr Institute, University of Copenhagen, 
Blegdamsvej 17, Copenhagen $\O$, Denmark 
\email{keiichi.nagao.phys@vc.ibaraki.ac.jp; hbech@nbi.dk}
}

%


\begin{abstract}%
We study the periodic complex action theory (CAT) 
by imposing a periodic condition 
in the future-included CAT where the time integration is performed 
from the past to the future, 
and extend a normalized matrix element of an operator $\hat{\mathcal O}$, 
which is called the weak value in the real action theory,  
to another expression $\langle \hat{\mathcal O} \rangle_{\mathrm{periodic}~\mathrm{time}}$. 
We present two theorems stating that 
$\langle \hat{\mathcal O} \rangle_{\mathrm{periodic}~\mathrm{time}}$ becomes real 
for $\hat{\mathcal O}$ being Hermitian 
with regard to a modified inner product 
that makes a given non-normal Hamiltonian $\hat{H}$ normal. 
The first theorem holds for a given period $t_p$ in a case where 
the number of eigenstates having the maximal imaginary part $B$ 
of the eigenvalues of $\hat{H}$ is just one, while 
the second one stands for $t_p$ selected such that  
the absolute value of the transition amplitude is maximized
in a case where $B \leq 0$ and $|B|$ is much smaller 
than the distances between any two real parts of the eigenvalues of  $\hat{H}$. 
The latter proven via a number-theoretical argument 
suggests that, if our universe is periodic, then even the period 
could be an adjustment parameter to be determined in the Feynman path integral. 
This is a variant type of the maximization principle that we previously proposed. 
\end{abstract}


\maketitle

%
\noindent 
{\bf\large  1. Introduction} \\
\noindent 
In the usual quantum theory  reality of action is implicitly imposed at first. 
Indeed, in the Feynman path integral, action is regarded as a phase of the integrand. 
However, there is a possibility that action also produces a scale factor in the integrand 
by taking a complex value.  
Such a complex action theory (CAT)~\cite{Bled2006} -- an attempt to describe quantum theory 
whose action is complex at a fundamental level but effectively looks real -- 
has been investigated intensively with the expectation that the imaginary part of the action 
would give some falsifiable predictions~\cite{Bled2006,Nielsen:2008cm,Nielsen:2007ak,Nielsen:2005ub}. 
Various interesting suggestions have been made for the Higgs mass~\cite{Nielsen:2007mj}, 
quantum-mechanical philosophy~\cite{newer1,Vaxjo2009,newer2}, 
some fine-tuning problems~\cite{Nielsen2010qq,degenerate}, 
black holes~\cite{Nielsen2009hq}, 
de Broglie--Bohm particles and a cut-off in loop diagrams~\cite{Bled2010B}, 
a mechanism to obtain Hermitian Hamiltonians~\cite{Nagao:2010xu}, 
the complex coordinate formalism~\cite{Nagao:2011za}, 
and the momentum relation~\cite{Nagao:2011is,Nagao:2013eda}. 
The CAT is classified into two types. 
One is a special type of theory that we call ``future-included". 
In the future-included theory, not only the past state $| A(T_A) \rangle$ at the initial time $T_A$ 
but also the future state $| B(T_B) \rangle$ at the final time $T_B$ is given at first, 
and the time integration is performed over the whole period 
from the past to the future. 
This is in contrast to the other usual type of theory that we call ``future-not-included", where 
only the past state $| A(T_A) \rangle$ is given at first, 
and the time integration is performed over the period between the initial time $T_A$ and 
some specific time $t$ ($T_A \leq t \leq T_B$). 
In Ref.~\cite{Nagao:2013eda} 
we clarified various interesting properties of the future-not-included CAT. 
However, in Ref.~\cite{Nagao:2017ecx}, we argued that, if a theory is described 
with a complex action, then such a theory is suggested to be the future-included theory, 
rather than the future-not-included theory. We encounter a philosophical 
contradiction in the future-not-included CAT as long as we respect objectivity.

In the future-included theory, what is expected to work as an expectation value 
for an operator $\hat{\mathcal O}$ is the normalized matrix element~\cite{Bled2006}\footnote{The normalized matrix element $\langle \hat{\mathcal O} \rangle^{BA}$ is called 
the weak value~\cite{AAV} in the context of the real action theory (RAT), 
and it has been intensively studied. For details, see Ref.~\cite{review_wv} and references therein.}  
$\langle \hat{\mathcal O} \rangle^{BA} 
\equiv \frac{ \langle B(t) |  \hat{\mathcal O}  | A(t) \rangle }{ \langle B(t) | A(t) \rangle }$. 
Indeed, if we regard $\langle \hat{\mathcal O} \rangle^{BA}$ 
as an expectation value in the future-included theory, 
we obtain the Heisenberg equation, Ehrenfest's theorem, 
and a conserved probability current density~\cite{Nagao:2012mj,Nagao:2012ye}. 
Thus $\langle \hat{\mathcal O} \rangle^{BA}$ has very nice properties. 
Here we note that $\langle \hat{\mathcal O} \rangle^{BA}$ is generically complex 
even for Hermitian $\hat{\mathcal O}$ by its definition. 
On the other hand, if $\langle \hat{\mathcal O} \rangle^{BA}$ is desired to be 
an expectation value for $\hat{\mathcal O}$, it has to be real, since we know that 
any observables should be real. 
Then how can we resolve this crucial problem?

In the CAT the imaginary parts of the eigenvalues $\lambda_i$
of a given non-normal Hamiltonian\footnote{The Hamiltonian $\hat{H}$ is 
generically non-normal, so it is 
not restricted to the class of PT-symmetric non-Hermitian Hamiltonians that were 
studied in 
Refs.~\cite{Bender:1998ke,Bender:1998gh,Mostafazadeh_CPT_ip_2002,Mostafazadeh_CPT_ip_2003,Bender:2011ke}.} $\hat{H}$ are supposed to be bounded from above for 
the Feynman path integral $\int e^{\frac{i}{\hbar}S} {\mathcal D} \text{path}$ to converge. 
We can imagine that some $\text{Im} \lambda_i$ take the maximal value $B$. 
We denote the corresponding subset of $\{ i \}$ as $A$. 
In Refs.~\cite{Nagao:2015bya,Nagao:2017cpl}, 
under this supposition, we answered the above question by proposing a theorem that states that, 
provided that an operator $\hat{\mathcal O}$ is $Q$-Hermitian, i.e., 
Hermitian with regard to a modified inner product $I_Q$ that makes the given Hamiltonian normal 
by using an appropriately chosen Hermitian operator $Q$, 
the normalized matrix element defined with $I_Q$ becomes real and 
time-develops under a $Q$-Hermitian Hamiltonian for the past and future states selected 
such that the absolute value of the transition amplitude defined with $I_Q$ from the past state 
to the future state is maximized. 
We call this way of thinking the maximization principle. 
In Ref.~\cite{Nagao:2015bya} we gave a proof of the theorem in the case of non-normal 
Hamiltonians $\hat{H}$ by considering that 
essentially only terms associated with the largest imaginary parts 
of the eigenvalues of $\hat{H}$, which belong to the subset $A$, 
contribute most significantly to the absolute value of the transition amplitude defined with $I_Q$, 
and that the normalized matrix element defined with $I_Q$ 
for such maximizing states becomes an expression similar to 
an expectation value defined with $I_Q$ in the future-not-included theory. 
This proof is based on the existence of imaginary parts of the eigenvalues of $\hat{H}$. 
In the case of the RAT we gave another proof in Ref.~\cite{Nagao:2017cpl}. 
In Ref.~\cite{Nagao:2015bya} we found that 
via the maximization principle 
in the expansion of the resulting maximizing states $| A(T_A) \rangle_{\rm{max}}$ and 
$| B(T_B) \rangle_{\rm{max}}$ in terms of the eigenstates of $\hat{H}$, 
$|A (T_A) \rangle_{\rm{max}} = \sum_{i \in A} a_i (T_A) | \lambda_i \rangle$, 
$|B (T_B) \rangle_{\rm{max}} = \sum_{i \in A} b_i (T_B) | \lambda_i \rangle$,  
the absolute values of each component were found to be the same: 
$|a_i (T_A)| = |b_i (T_B)|~ \text{for $\forall i \in A$}$, while the phases were not so. 
This fact has partly motivated us to study a periodic universe. 
In addition, it would be interesting by itself to look for a possibility that our universe runs periodically, 
and also to see whether there still exists any kind of reality theorems on 
the expectation value for $\hat{\mathcal O}$ in such a periodic CAT.

In this letter, 
after briefly reviewing the future-included CAT and maximization principle, 
we study the periodic CAT. 
For simplicity let us now suppose that we obtained a periodic universe via the maximization principle 
for the past and future states $| A(T_A) \rangle$ and $| B(T_B) \rangle$, or just consider it  
by imposing a periodic condition on the past and future states in the future-included CAT at first. 
Then the remaining quantity to be adjusted could be 
the period. If so, it would be very interesting that even the period is regarded 
as a parameter that is adjusted via the maximization principle. 
To proceed with this speculation, taking $T_B=T_A + t_p$, we simply impose 
the following periodic condition\footnote{Another periodic condition might be 
$ | A(T_A) \rangle =  | A(T_A + t_p) \rangle = e^{-\frac{i}{\hbar} \hat{H} t_p } | A(T_A) \rangle$. 
This means that $e^{-\frac{i}{\hbar} \hat{H} t_p }$ has to have $| A(T_A) \rangle$ 
as the eigenstate for its eigenvalue $1$. 
Since $| A(T_A) \rangle$ is supposed to be a generic state,  
we do not adopt such a periodic condition in this letter. } 
on the future and past states in the future-included CAT: 
\begin{eqnarray}
| B(T_B) \rangle = | B(T_A + t_p) \rangle =  | A(T_A) \rangle .    \label{periodic_condition_on_B_and_A}  
\end{eqnarray}
In the periodic CAT, 
extending a normalized matrix element of an operator $\hat{\mathcal O}$ 
to an expression such that various normalized matrix elements of $\hat{\mathcal O}$ 
are summed up with the weight of transition amplitudes,  
we introduce another normalized quantity 
$\langle \hat{\mathcal O} \rangle_{\mathrm{periodic}~\mathrm{time}} 
\equiv 
\frac{\mathrm{Tr} \left(   e^{-\frac{i}{\hbar} \hat{H} t_p }  \hat{\mathcal O} \right) }{ \mathrm{Tr}  \left(   e^{-\frac{i}{\hbar} \hat{H} t_p}  \right) }$ 
that is generically complex but expected to have a role of an expectation value for $\hat{\mathcal O}$. 
We present two theorems stating that 
$\langle \hat{\mathcal O} \rangle_{\mathrm{periodic}~\mathrm{time}}$ becomes real 
for $Q$-Hermitian $\hat{\mathcal O}$. 
The first theorem holds for a given period $t_p$, 
even without any adjustment of it, 
in a case where the order of the subset $A$ is just one, i.e., 
the number of eigenstates having the maximal imaginary part $B$ 
of the eigenvalues of $\hat{H}$ is just one, 
while the second one stands for $t_p$ selected such that 
the absolute value of the transition amplitude 
$|\mathrm{Tr}  \left(   e^{-\frac{i}{\hbar} \hat{H} t_p}  \right)|$ is maximized 
in a case where $B \leq 0$ and $|B|$ is much smaller 
than the distances between any two real parts of the eigenvalues of  $\hat{H}$. 
The second theorem, which is proven partly via a number-theoretical argument, 
suggests that, if our universe is periodic, then even the period 
could be an adjustment parameter to be determined 
in the Feynman path integral 
via such a variant type of the maximization principle that we proposed 
in Refs.~\cite{Nagao:2015bya,Nagao:2017cpl}.

\vspace*{10mm}
%
\noindent 
{\bf\large  2. Future-included complex action theory and maximization principle} \\
\noindent 
The eigenstates of a given non-normal Hamiltonian 
$\hat{H}$, 
$| \lambda_i \rangle (i=1,2,\dots)$ 
obeying $\hat{H} | \lambda_i \rangle = \lambda_i | \lambda_i \rangle$,  are not orthogonal to 
each other in the usual inner product $I$. 
In order to obtain an orthogonal basis, let us introduce a modified inner product 
$I_Q$~\cite{Nagao:2010xu,Nagao:2011za} 
that makes $\hat{H}$ normal with respect to it.\footnote{Similar inner products are also studied 
in Refs.~\cite{Geyer,Mostafazadeh_CPT_ip_2002,Mostafazadeh_CPT_ip_2003}.}  
This enables $| \lambda_i \rangle (i=1,2,\dots)$ to be orthogonal to each other 
with regard to $I_Q$, which is defined 
for arbitrary kets $|u \rangle$ and $|v \rangle$ as 
$I_Q(|u \rangle , |v \rangle) \equiv \langle u |_Q v \rangle 
\equiv \langle u | Q | v \rangle$. 
Here $Q$ is a Hermitian operator that obeys $\langle \lambda_i  |_Q \lambda_j \rangle = \delta_{ij}$. 
The Hamiltonian $\hat{H}$ is diagonalized as 
$\hat{H} = PD P^{-1}$, where
$P=(| \lambda_1 \rangle , | \lambda_2 \rangle , \ldots)$ and 
$D=\text{diag}(\lambda_1, \lambda_2, \dots)$.  
Using the diagonalizing operator $P$, 
we choose $Q=(P^\dag)^{-1} P^{-1}$. 
Utilizing this $Q$, we introduce the $Q$-Hermitian conjugate $\dag^Q$ of 
an operator $A$ by $\langle \psi_2 |_Q A | \psi_1 \rangle^* \equiv \langle \psi_1 |_Q A^{\dag^Q} | \psi_2 \rangle$, so $A^{\dag^Q} \equiv Q^{-1} A^\dag Q$. 
Also, we define $\dag^Q$ for kets and bras as 
$| \lambda \rangle^{\dag^Q} \equiv \langle \lambda |_Q $ and 
$\left(\langle \lambda |_Q \right)^{\dag^Q} \equiv | \lambda \rangle$. 
If $A$ obeys $A^{\dag^Q} = A$, we call $A$ $Q$-Hermitian. 
We note that, since 
$P^{-1}=
\left(
 \begin{array}{c}
      \langle \lambda_1 |_Q     \\
      \langle \lambda_2 |_Q     \\
      \vdots 
 \end{array}
\right)$ 
obeys $P^{-1} \hat{H} P = D$ and 
$P^{-1} \hat{H}^{\dag^Q} P = D^{\dag}$, 
$\hat{H}$ is $Q$-normal, 
$[\hat{H}, \hat{H}^{\dag^Q} ] = P [D, D^\dag ] P^{-1} =0$. 
$\hat{H}$ can be decomposed as 
$\hat{H}=\hat{H}_{Qh} + \hat{H}_{Qa}$, 
where $\hat{H}_{Qh}= \frac{\hat{H} + \hat{H}^{\dag^Q} }{2}$ and 
$\hat{H}_{Qa} = \frac{\hat{H} - \hat{H}^{\dag^Q} }{2}$ are 
$Q$-Hermitian and anti-$Q$-Hermitian parts of $\hat{H}$, respectively.

In Ref.~\cite{Nagao:2015bya}, we adopted the modified inner product $I_Q$ 
for all quantities in the future-included CAT~\cite{Bled2006,Nagao:2012mj,Nagao:2012ye}. 
The future-included CAT is described by using 
the future state $| B (T_B) \rangle$ at the final time $T_B$ 
and the past state $| A (T_A) \rangle$ at the initial time $T_A$, 
where $| A (T_A) \rangle$ and $| B (T_B) \rangle$ are supposed to 
time-develop as follows: 
\begin{eqnarray}
&&i \hbar \frac{d}{dt} | A(t) \rangle = \hat{H} | A(t) \rangle , \label{schro_eq_Astate} \\
&&
i \hbar \frac{d}{dt} | B(t) \rangle = {\hat{H}}^{\dag^Q} | B(t) \rangle 
\quad \Leftrightarrow \quad 
-i \hbar \frac{d}{dt} \langle B(t) |_Q  
= \langle B(t) |_Q  \hat{H}. 
\label{schro_eq_Bstate} 
\end{eqnarray}
The normalized matrix element defined with the modified inner product $I_Q$ is expressed as 
\begin{equation}
\langle \hat{\mathcal O} \rangle_Q^{BA} 
\equiv \frac{ \langle B(t) |_Q  \hat{\mathcal O}  | A(t) \rangle }{ \langle B(t) |_Q A(t) \rangle }. 
\label{OQBA}
\end{equation} 
If we change the notation of $\langle B(t) |$ such that it absorbs $Q$, it can be expressed simply as 
$\langle \hat{\mathcal O} \rangle^{BA}$~\cite{Bled2006}. 
In the case of $Q=1$, this corresponds to the weak value~\cite{AAV, review_wv} 
that is well known in the RAT. 
If we regard $\langle \hat{\mathcal O} \rangle_Q^{BA}$ 
as an expectation value in the future-included CAT, 
then we obtain the Heisenberg equation, Ehrenfest's theorem, 
and a conserved probability current density~\cite{Nagao:2012mj,Nagao:2012ye}.  
Therefore, this quantity is a good candidate for an expectation value 
in the future-included CAT.

In Ref.~\cite{Nagao:2015bya} we proposed the following theorem.

\vspace{0.5cm}

\noindent 
{\bf Theorem 1.} 
{\em 
As a prerequisite, assume that a given Hamiltonian 
$\hat{H}$ is non-normal but diagonalizable 
and that the imaginary parts of the eigenvalues 
of $\hat{H}$ are bounded from above, 
and define a modified inner product $I_Q$ by means 
of a Hermitian operator $Q$ arranged so 
that $\hat{H}$ becomes normal with respect to $I_Q$. 
Let the two states $| A(t) \rangle$ and $ | B(t) \rangle$ 
time-develop according to the Schr\"{o}dinger equations 
with $\hat{H}$ and $\hat{H}^{\dag^Q}$ respectively: 
$|A (t) \rangle = e^{-\frac{i}{\hbar}\hat{H} (t-T_A) }| A(T_A) \rangle$, 
$|B (t) \rangle = e^{-\frac{i}{\hbar} {\hat{H}}^{\dag^Q} (t-T_B) } | B(T_B)\rangle$, 
and be normalized with $I_Q$ 
at the initial time $T_A$ and the final time $T_B$ respectively: 
$\langle A(T_A) |_{Q} A(T_A) \rangle = 1$,
$\langle B(T_B) |_{Q} B(T_B) \rangle = 1$. 
Next determine $|A(T_A) \rangle$ and $|B(T_B) \rangle$ so as to maximize 
the absolute value of the transition amplitude 
$|\langle B(t) |_Q A(t) \rangle|=|\langle B(T_B)|_Q \exp(-i\hat{H}(T_B-T_A)) |A(T_A) \rangle|$. 
Then, provided that an operator $\hat{\mathcal O}$ is $Q$-Hermitian, i.e., Hermitian 
with respect to the inner product $I_Q$, 
$\hat{\mathcal O}^{\dag^Q} = \hat{\mathcal O}$, 
the normalized matrix element of the operator $\hat{\mathcal O}$ defined by 
$\langle \hat{\mathcal O} \rangle_Q^{BA} 
\equiv
\frac{\langle B(t) |_Q \hat{\mathcal O} | A(t) \rangle}{\langle B(t) |_Q A(t) \rangle}$ 
becomes {\rm real} and time-develops under 
a {\rm $Q$-Hermitian} Hamiltonian. }


\vspace*{0.5cm}

We call this way of thinking the maximization principle. 
To prove this theorem in the case of the CAT\footnote{The above proof 
depends on the existence of 
imaginary parts of the eigenvalues of $\hat{H}$, so it does not apply to 
the case where a given Hamiltonian is Hermitian, 
where there are no imaginary parts of the eigenvalues. 
For the proof in such a special case, see Ref.~\cite{Nagao:2017cpl}.  
The maximization principle is reviewed in 
Refs.~\cite{Nagao:2017book,Nagao:2017ztx}. }, 
we expand $| A(t) \rangle$ and $| B(t) \rangle$ in terms of the eigenstates $| \lambda_i \rangle$ 
as follows: 
$|A (t) \rangle = \sum_i a_i (t) | \lambda_i \rangle$,  
$|B (t) \rangle = \sum_i b_i (t) | \lambda_i \rangle$, 
where 
$a_i (t) = a_i (T_A) e^{-\frac{i}{\hbar}\lambda_i (t-T_A) }$, 
$b_i (t) = b_i (T_B) e^{-\frac{i}{\hbar}\lambda_i^* (t-T_B) }$. 
Let us express $a_i(T_A)$ and $b_i(T_B)$ as 
$a_i(T_A)= | a_i(T_A) | e^{i \theta_{a_i}}$ and 
$b_i(T_B) = | b_i(T_B) | e^{i \theta_{b_i}}$, and introduce 
$T\equiv T_B - T_A$ and $\Theta_i \equiv \theta_{a_i} - \theta_{b_i} 
- \frac{1}{\hbar} T \text{Re} \lambda_i$. 
Since the imaginary parts of the eigenvalues 
of $\hat{H}$ are supposed to be bounded from above 
for the Feynman path integral
$\int e^{\frac{i}{\hbar}S} {\mathcal D} \text{path}$ 
to converge, we can imagine that some of $\text{Im} \lambda_i$ 
take the maximal value $B$, and denote the corresponding subset of $\{ i \}$ as $A$. 
Then, $| \langle B (t) |_Q A (t) \rangle |$ can take 
the maximal value $e^{\frac{1}{\hbar} B T}$ only under the following conditions: 
$\Theta_i  \equiv \Theta_c \quad \text{for $\forall i \in A$}$, 
$\sum_{i \in A} | a_i (T_A) |^2 =\sum_{i \in A}|b_i (T_B)|^2 = 1$, 
$|a_i (T_A)| = |b_i (T_B)|  \quad \text{for $\forall i \in A$}$, 
$| a_i (T_A) |  = | b_i (T_B) | =0 \quad \text{for $\forall i \notin A$}$, 
and the states to maximize $| \langle B (t) |_Q A (t) \rangle |$ are expressed as 
$|A (t) \rangle_{\rm{max}} = \sum_{i \in A} a_i (t) | \lambda_i \rangle$ and 
$|B (t) \rangle_{\rm{max}} = \sum_{i \in A} b_i (t) | \lambda_i \rangle$. 
Introducing $| \tilde{A}(t) \rangle \equiv 
e^{-\frac{i}{\hbar}(t-T_A) \hat{H}_{Qh}} $ $| A(T_A) \rangle_{\rm{max}}$, 
which is normalized as 
$\langle \tilde{A}(t) |_Q \tilde{A}(t) \rangle = 1$ 
and obeys the Schr\"{o}dinger equation 
$i\hbar  \frac{d}{d t}| \tilde{A}(t) \rangle = \hat{H}_{Qh} | \tilde{A}(t) \rangle$, 
the normalized matrix element 
for $| A(t) \rangle_{\rm{max}}$ and $| B(t) \rangle_{\rm{max}}$ is evaluated as 
$\langle \hat{\mathcal O} \rangle_Q^{B_{\rm{max}} A_{\rm{max}}} =
\langle \tilde{A}(t) |_Q \hat{\mathcal O} | \tilde{A}(t) \rangle$. 
Hence $\langle \hat{\mathcal O} \rangle_Q^{B_{\rm{max}} A_{\rm{max}}}$ is real 
for $Q$-Hermitian $\hat{\mathcal O}$, and time-develops 
under the $Q$-Hermitian Hamiltonian $\hat{H}_{Qh}$: 
$\frac{d}{dt} \langle \hat{\mathcal O} \rangle_Q^{B_{\rm{max}} A_{\rm{max}}}
=\frac{i}{\hbar} \langle \left[ \hat{H}_{Qh}, \hat{\mathcal O} \right] \rangle_Q^{B_{\rm{max}} A_{\rm{max}}}$. 
Thus we have seen that the maximization principle provides both 
the reality of $\langle \hat{\mathcal O} \rangle_Q^{BA}$ 
for $Q$-Hermitian $\hat{\mathcal O}$ and the $Q$-Hermitian Hamiltonian.

\vspace*{10mm}
%
%
\noindent 
{\bf\large  3. Periodic complex action theory and maximization principle} \\
\noindent 
In the future-included CAT, let us take $T_B=T_A + t_p$, and impose the periodicity condition 
(\ref{periodic_condition_on_B_and_A}). 
Then, since $\langle B(t) |_Q$ is expressed as 
$\langle B(t) |_Q =\langle  A(T_A) |_Q  e^{\frac{i}{\hbar} \hat{H}(t - T_B) }$, 
$\langle \hat{\mathcal O} \rangle^{BA}_Q$
defined in Eq.(\ref{OQBA}) is written as 
\begin{equation}
\langle \hat{\mathcal O} \rangle^{BA}_Q 
=
\frac{ \langle  A(T_A) |_Q  e^{\frac{i}{\hbar} \hat{H}(t - T_B) }  \hat{\mathcal O}  e^{-\frac{i}{\hbar} \hat{H} (t - T_A) } | A(T_A) \rangle }{\langle  A(T_A) |_Q e^{-\frac{i}{\hbar} \hat{H} t_p} | A(T_A) \rangle } 
\equiv
\langle \hat{\mathcal O} \rangle_{Q}^A. 
\label{OBA_for_periodic_case}
\end{equation} 
Now we rewrite $\langle \hat{\mathcal O} \rangle_{Q}^A$ as follows: 
\begin{equation}
\langle \hat{\mathcal O} \rangle_{Q}^A 
\simeq
\frac{\sum_n 
\langle \hat{\mathcal O} \rangle_Q^{A_n}~ \langle  A_n |_Q e^{-\frac{i}{\hbar} \hat{H} t_p} | A_n \rangle}{\sum_n \langle  A_n |_Q e^{-\frac{i}{\hbar} \hat{H} t_p} | A_n \rangle } 
=
\frac{\mathrm{Tr} \left(   e^{-\frac{i}{\hbar} \hat{H} t_p }  \hat{\mathcal O} \right) }{ \mathrm{Tr}  \left(   e^{-\frac{i}{\hbar} \hat{H} t_p}  \right) } , 
\label{O_A_O_periodic}
\end{equation} 
where we have taken a basis $| A_n \rangle$ such that the state $| A(T_A) \rangle$ maximizing 
$|\langle  A(T_A) |_Q e^{-\frac{i}{\hbar} \hat{H} t_p} | A(T_A) \rangle|$ is included. 
Weighting various normalized matrix elements 
$\langle \hat{\mathcal O} \rangle_Q^{A_n}$ by 
$\langle  A_n |_Q e^{-\frac{i}{\hbar} \hat{H} t_p} | A_n \rangle$ 
replaces maximizing $|\langle  A(T_A) |_Q e^{-\frac{i}{\hbar} \hat{H} t_p} | A(T_A) \rangle|$ 
crudely in a quantitative way. 
In addition we have used the cyclic property of $\mathrm{Tr}$.

Based on the above evaluation, in the periodic CAT specified by the periodic condition (\ref{periodic_condition_on_B_and_A}), 
we propose our ``expectation value" for $\hat{\mathcal O}$ by 
the following quantity:  
\begin{eqnarray}
\langle \hat{\mathcal O} \rangle_{\mathrm{periodic}~\mathrm{time}} 
&\equiv& 
\frac{\mathrm{Tr} \left(   e^{-\frac{i}{\hbar} \hat{H} t_p }  \hat{\mathcal O} \right) }{ \mathrm{Tr}  \left(   e^{-\frac{i}{\hbar} \hat{H} t_p}  \right) }.  \label{Ohat_periodic}
\end{eqnarray} 
This quantity is generically complex by its definition, so it is unclear whether we can use it 
as an expectation value for $\hat{\mathcal O}$. 
In addition, this quantity is independent of the time $t$, so the situation is like that in general relativity 
with an exact symmetry under translations in the time variable, where there is 
conservation of the total energy, which is even just zero, 
and averaging would lead to no time dependence.
If we want to reintroduce the time $t$ dependence, as we are accustomed to, we would have 
to introduce a clock variable $T_\mathrm{clock}(t)$ to be inserted in the normalized quantity 
of Eq.(\ref{Ohat_periodic}). 
In this letter, however, we will not be involved in it, but we concentrate on whether 
$\langle \hat{\mathcal O} \rangle_{\mathrm{periodic}~\mathrm{time}}$ could be real, since 
the reality of $\langle \hat{\mathcal O} \rangle_{\mathrm{periodic}~\mathrm{time}}$ is 
crucially important 
for our theory to be viable. 
Seeking a condition for $\langle \hat{\mathcal O} \rangle_{\mathrm{periodic}~\mathrm{time}}$ to be real 
provided that $\hat{\mathcal O}$ is $Q$-Hemitian,  
we propose the following two theorems in special cases. 
In the first theorem, we consider a case where the order of 
the subset\footnote{The subset $A$ is given in the proof of Theorem 1. 
$B$ is the maximal value of $\mathrm{Im} \lambda_n$. } $A$ is just one for the given period $t_p$. 
In the second theorem, 
assuming that the maximal value $B$ of the imaginary parts of 
the eigenvalues of $\hat{H}$ is equal to or smaller than 0, and that $|B|$ is much smaller 
than the distances between any two real parts of the eigenvalues of  $\hat{H}$, 
we regard $t_p$ as an adjustment parameter, which is to be selected 
such that the absolute value of the transition amplitude 
$|\mathrm{Tr}  \left(   e^{-\frac{i}{\hbar} \hat{H} t_p}  \right)|$ is maximized. 
The second theorem is a variant type of Theorem 1 
in the point that, on behalf of $| B(T_B) \rangle$ and $| A(T_A) \rangle$ that are 
constrained by the condition (\ref{periodic_condition_on_B_and_A}), the period 
$t_p$ is used as an adjustment parameter.


\vspace{0.5cm}

\noindent 
{\bf Theorem 2.} 
{\em 
As a prerequisite, assume that a given Hamiltonian 
$\hat{H}$ is non-normal but diagonalizable 
and 
that only one significant eigenstate of  $\hat{H}$ contributes essentially for a given fixed period $t_p$, 
and define a modified inner product $I_Q$ by means 
of a Hermitian operator $Q$ arranged so 
that $\hat{H}$ becomes normal with respect to $I_Q$. 
Then, provided that an operator $\hat{\mathcal O}$ is $Q$-Hermitian, i.e., Hermitian 
with respect to the inner product $I_Q$, 
$\hat{\mathcal O}^{\dag^Q} = \hat{\mathcal O}$, 
$\langle \hat{\mathcal O} \rangle_{\mathrm{periodic}~\mathrm{time}} \equiv \frac{\mathrm{Tr} \left(   e^{-\frac{i}{\hbar} \hat{H} t_p }  \hat{\mathcal O} \right) }{ \mathrm{Tr}  \left(   e^{-\frac{i}{\hbar} \hat{H} t_p}  \right) }$ 
becomes real. }  

\vspace*{0.5cm}

\noindent 
{\bf Theorem 3.} 
{\em 
As a prerequisite, assume that a given Hamiltonian 
$\hat{H}$ is non-normal but diagonalizable, 
that the maximal value $B$ of the imaginary parts of the eigenvalues 
of $\hat{H}$ is equal to or smaller than zero, and  
that $|B|$ is much smaller than 
the distances between any two real parts of the eigenvalues of $\hat{H}$, 
and define a modified inner product $I_Q$ by means 
of a Hermitian operator $Q$ arranged so 
that $\hat{H}$ becomes normal with respect to $I_Q$. 
Then, provided that an operator $\hat{\mathcal O}$ is $Q$-Hermitian, i.e., Hermitian 
with respect to the inner product $I_Q$, 
$\hat{\mathcal O}^{\dag^Q} = \hat{\mathcal O}$, 
$\langle \hat{\mathcal O} \rangle_{\mathrm{periodic}~\mathrm{time}} \equiv \frac{\mathrm{Tr} \left(   e^{-\frac{i}{\hbar} \hat{H} t_p }  \hat{\mathcal O} \right) }{ \mathrm{Tr}  \left(   e^{-\frac{i}{\hbar} \hat{H} t_p}  \right) }$ 
becomes real for selected periods $t_p$ 
such that $|\mathrm{Tr}  \left(   e^{-\frac{i}{\hbar} \hat{H} t_p}  \right)|$ is maximized. }


\vspace*{0.5cm}

In preparation for 
proving these theorems, we first evaluate the numerator and denominator of the right-hand side of 
Eq.(\ref{Ohat_periodic}). The numerator is expressed as 
$\mathrm{Tr}  \left(   e^{-\frac{i}{\hbar} \hat{H} t_p}  \hat{\mathcal O}  \right)
= 
\sum_n \langle  \lambda_n |_Q  e^{-\frac{i}{\hbar} \hat{H} t_p} \hat{\mathcal O} |  \lambda_n \rangle 
\simeq 
e^{\frac{B}{\hbar} t_p} 
\sum_{n \in A} \langle  \lambda_n |_Q   \hat{\mathcal O} |  \lambda_n \rangle e^{-i \theta_n }$, 
where we have used as a basis the set of eigenstates of the Hamiltonian $\hat{H}$, 
$| \lambda_n \rangle$, 
which obeys the orthogonality and completeness relations: $\langle  \lambda_n |_Q  \lambda_m \rangle=\delta_{nm}$, 
$\sum_m | \lambda_m \rangle \langle \lambda_m |_Q =1$. 
In addition, we have introduced  
$\theta_n \equiv \frac{1}{\hbar} \mathrm{Re} \lambda_n t_p$,  
and supposed that $t_p$ is sufficiently large from a phenomenological point of view 
so that the terms coming from the subset $A$ dominate most significantly. 
Similarly the denominator is evaluated as 
$\mathrm{Tr}  \left(   e^{-\frac{i}{\hbar} \hat{H} t_p}    \right)
\simeq 
e^{\frac{B}{\hbar} t_p} 
\sum_{n \in A}  e^{- i \theta_n}$. 
Thus $\langle \hat{\mathcal O} \rangle_{\mathrm{periodic}~\mathrm{time}}$ 
is reduced to the following expression: 
\begin{eqnarray}
\langle \hat{\mathcal O} \rangle_{\mathrm{periodic}~\mathrm{time}} 
&\simeq& \frac{\sum_{n \in A} \langle  \lambda_n |_Q   \hat{\mathcal O} |  \lambda_n \rangle e^{-i \theta_n } }{ \sum_{n \in A}  e^{-i \theta_n }} . 
\label{Ohat_periodic_new_A}
\end{eqnarray}

First let us prove Theorem 2 
for a given fixed period $t_p$ 
by assuming that the order of the subset $A$ is one. 
We express the dominating eigenstate and 
eigenvalue associated with it 
as $|\lambda_d \rangle$ and $\lambda_d$ respectively. 
Then, since both the numerator and denominator of the right-hand side of 
Eq.(\ref{Ohat_periodic_new_A}) are composed of only one term 
associated with $\lambda_d$, 
$\langle \hat{\mathcal O} \rangle_{\mathrm{periodic}~\mathrm{time}}$ 
is expressed as 
\begin{eqnarray}
\langle \hat{\mathcal O} \rangle_{\mathrm{periodic}~\mathrm{time}} 
&\simeq& 
\langle \lambda_d |_Q \hat{\mathcal O} | \lambda_d \rangle. 
\label{Ohat_periodic_new_order_of_A_is_1}
\end{eqnarray}
This is real for $Q$-Hermitian $\hat{\mathcal O}$, so we have proven Theorem 2.

Next let us prove Theorem 3 by utilizing the expression of Eq.(\ref{Ohat_periodic_new_A}) again. 
The function $f(t_p) \equiv \lvert \mathrm{Tr}  \left(   e^{-\frac{i}{\hbar} \hat{H} t_p}  \right) \rvert^2$ 
and its derivative with regard to $t_p$ are evaluated as 
$f(t_p) 
\simeq
e^{\frac{2}{\hbar} B t_p} 
\sum_{n,m \in A} 
\cos\left\{ \frac{1}{\hbar} ( \mathrm{Re} \lambda_m - \mathrm{Re} \lambda_n ) t_p \right\}$ 
and  
$\frac{df(t_p)}{dt_p} 
\simeq
\frac{1}{\hbar} \sum_{n,m \in A} 
\left[ 
2B \cos\left\{ \frac{1}{\hbar} ( \mathrm{Re} \lambda_m - \mathrm{Re} \lambda_n ) t_p \right\} 
\right.$ 
$\left. - \sin\left\{ \frac{1}{\hbar} ( \mathrm{Re} \lambda_m - \mathrm{Re} \lambda_n ) t_p \right\} 
 ( \mathrm{Re} \lambda_m - \mathrm{Re} \lambda_n ) \right]  
e^{\frac{2B}{\hbar} t_p}$. 
Since we are assuming\footnote{We note that $B \le 0$ has to be supposed so that 
$\lvert \mathrm{Tr}  \left(   e^{-\frac{i}{\hbar} \hat{H} t_p}  \right) \rvert = 
e^{\frac{B}{\hbar} t_p} \lvert \sum_{n \in A}  e^{- i \theta_n} \rvert$ 
does not diverge when 
we seek $t_p$ such that $\lvert \mathrm{Tr}  \left(   e^{-\frac{i}{\hbar} \hat{H} t_p}  \right) \rvert$ 
is maximized. } 
that $B \le 0$ and $|B|$ is much smaller than the distances 
between any two real parts of the eigenvalues of  $\hat{H}$, the second term in 
the square brackets contributes significantly in 
the expression of $\frac{df(t_p)}{dt_p}$. 
Thus we find that, for $\theta_i$ such that  
\begin{equation}
\theta_i = \theta_c ~(\mathrm{mod}~ 2 \pi) ~\mathrm{for}~ \forall  i \in A 
\Leftrightarrow
\mathrm{Re} \lambda_i t_p = \hbar \theta_c \equiv C  
~(\mathrm{mod}~ 2 \pi \hbar) ~\mathrm{for}~ \forall  i  \in A ,  
\label{cond_Relambdait_p=hbarthetac2} 
\end{equation}
$\frac{d^2 f(t_p)}{d {t_p}^2}<0$ and $f(t_p)$ is maximized. 
In Eq.(\ref{cond_Relambdait_p=hbarthetac2}) we have introduced $C \equiv \hbar \theta_c$. 
If $t_p$ satisfying Eq.(\ref{cond_Relambdait_p=hbarthetac2}) exist,  
the phase factor $e^{- i \theta_n}$ 
becomes the same for $\forall n \in A$ in Eq.(\ref{Ohat_periodic_new_A}), 
so $\langle \hat{\mathcal O} \rangle_{\mathrm{periodic} ~\mathrm{time}}$ 
is reduced to a simpler expression: 
\begin{eqnarray}
\langle \hat{\mathcal O} \rangle_{\mathrm{periodic}~\mathrm{time}} 
&\simeq& \frac{\sum_{n \in A} \langle \lambda_n |_Q \hat{\mathcal O} | \lambda_n \rangle }{ \sum_{n \in A} 1 }. 
\label{Ohat_periodic_new}
\end{eqnarray}
This is real for $Q$-Hermitian $\hat{\mathcal O}$. 
Thus Theorem 3 will be proven.


\vspace*{10mm}
%
\noindent 
{\bf\large  4. Proof of the existence of $t_p$ satisfying Eq.(\ref{cond_Relambdait_p=hbarthetac2})} \\
\noindent 
The existence of $t_p$ satisfying Eq.(\ref{cond_Relambdait_p=hbarthetac2}) 
looks believable. 
Now we investigate it explicitly according to the order of the Hilbert space that is labeled by the subset $A$. 
First let us consider the case where the order of the Hilbert space is two. 
We express 
$\mathrm{Re} \lambda_i~(i \in A)$ as 
$\left\{ \mathrm{Re} \lambda_i \right\} = \left\{ \alpha_1, \alpha_2 \right\}$, 
where $\alpha_1 < \alpha_2$. 
In this case the condition (\ref{cond_Relambdait_p=hbarthetac2}) is expressed as 
$\alpha_1 t_p = C  + h m_1$ and $\alpha_2 t_p = C  + h m_2$, 
where $m_1$ and $m_2$ $(m_1 < m_2)$ are integers that are to be chosen properly,  
and $C$ is a constant $(0 \le C < h)$. 
In order for $t_p$ to obey these relations, 
there have to exist integers $m_1$ and $m_2$ that obey 
$\alpha_1 t_p - h m_1 = \alpha_2 t_p - h m_2 = C$
$\Leftrightarrow$ 
$t_p=\frac{h (m_2 - m_1)}{\alpha_2 -  \alpha_1}$ 
and $\frac{\alpha_2}{\alpha_1}=\frac{C+h m_2}{C+h m_1}$ leading to 
$C=\frac{h}{\alpha_2 - \alpha_1}(\alpha_1 m_2 - \alpha_2 m_1)$. 
The condition $0 \le C < h$ is expressed as 
$0\le\frac{1}{\alpha_2 - \alpha_1}(\alpha_1 m_2 - \alpha_2 m_1) < 1$, i.e., 
$\alpha_2 m_1 \le \alpha_1 m_2 < \alpha_2 m_1 + (\alpha_2 - \alpha_1)$, 
which allows many pairs of $(m_1, m_2)$. 
Thus, in the $B=0$ case, we obtain 
many $t_p$, for which 
$f(t_p)$ is maximized. 
In the very small $|B| \neq 0$ case, because of the factor $e^{\frac{2B}{\hbar} t_p}$ 
included in $f(t_p)$, 
most of such $t_p$ become values for local maxima, and 
only the smallest $t_p$, i.e., $t_p$ for $(m_1,m_2)$ giving the smallest $m_2-m_1$, 
is selected.

In the case where the order of the Hilbert space is three, we express  
$\mathrm{Re} \lambda_i ~(i \in A)$ as 
$\left\{ \mathrm{Re} \lambda_i \right\} =  \left\{ \alpha_1, \alpha_2, \alpha_3 \right\}$, 
where we suppose that $\alpha_1 < \alpha_2 < \alpha_3$. 
In this case the condition (\ref{cond_Relambdait_p=hbarthetac2}) is expressed as 
$\alpha_1 t_p = C  +h m_1$, $\alpha_2 t_p = C  +h m_2$, and $\alpha_3 t_p = C  +h m_3$, 
where $m_1$, $m_2$, and $m_3$ $(m_1 < m_2 < m_3)$ are integers that are to be chosen properly, 
and $C$ is a constant $(0 \le C < h)$. 
In order for $t_p$ to obey these relations, 
there have to exist integers $m_1$, $m_2$, and $m_3$ that obey 
$\alpha_1 t_p - h m_1 = \alpha_2 t_p - h m_2 = \alpha_3 t_p - h m_3 = C$
$\Leftrightarrow$ 
$t_p=\frac{h (m_2 - m_1)}{\alpha_2 -  \alpha_1}=\frac{h (m_3 - m_2)}{\alpha_3 -  \alpha_2}$ 
and $\frac{\alpha_2}{\alpha_1}=\frac{C+h m_2}{C+h m_1}$, which leads to 
$C=\frac{h}{\alpha_2 - \alpha_1}(\alpha_1 m_2 - \alpha_2 m_1)$.  
Let us suppose that the ratio of $\alpha_2 -  \alpha_1$ to $\alpha_3 -  \alpha_2$ and that 
of $\alpha_3 -  \alpha_2$ to $\alpha_3 -  \alpha_1$ are rational 
numbers\footnote{In the case where both of them are not rational numbers, i.e., 
incommensurable, we approximate the irrational numbers 
to rational ones in their neighborhoods. }, and  
express them as 
$\frac{\alpha_2 -  \alpha_1}{\alpha_3 -  \alpha_2} =\frac{m_2 - m_1}{m_3 - m_2} = \frac{n_1}{d_1}$ and 
$\frac{\alpha_3 -  \alpha_2}{\alpha_3 -  \alpha_1} =\frac{m_3 - m_2}{m_3 - m_1} = \frac{n_2}{d_2}$, 
where $n_i$ and $d_i$ $(i=1,2)$ are positive and co-prime integers.\footnote{We note that 
$\mathrm{gcd}(n_i, d_i) =1$ for $i=1,2$, where $\mathrm{gcd}(a, b)$ is the greatest common divisor 
of integers $a$ and $b$.}  
Since we have the relations $(m_3-m_2)n_1 = (m_2-m_1)d_1$ and $(m_3-m_1)n_2 = (m_3-m_2)d_2$, 
we find 
$(m_3 -m_2, m_2 -m_1)= k (d_1, n_1)$ and $(m_3 -m_1, m_3 - m_2)= l (d_2, n_2)$, where $k$ and $l$ 
are positive integers to be chosen properly. 
Then, we are led to the relation $m_3-m_2=k d_1 = l n_2$, so we find that $k$ and $l$ are expressed as 
$(k, l)=a(n_2/\mathrm{gcd}(n_2, d_1), d_1/ \mathrm{gcd}(n_2, d_1) )$, where $a$ is a positive integer 
to be chosen properly. Thus we obtain $m_3-m_2=a n_2 d_1/\mathrm{gcd}(n_2, d_1)$, 
$m_2-m_1=a n_1 n_2 /\mathrm{gcd}(n_2, d_1)$, and $m_3-m_1 = a d_1 d_2/\mathrm{gcd}(n_2, d_1)$.  
Since the first and second relations provide $m_3-m_1 = a n_2 (d_1 +n_1)/\mathrm{gcd}(n_2, d_1)$, 
comparing this with the third relation, we obtain the relation $d_1 d_2 = n_2 (d_1 + n_1)$, which leads 
to $n_2=d_1/\mathrm{gcd}(d_1, n_1+d_1)$ and $d_2= (n_1+d_1)/\mathrm{gcd}(d_1, n_1+d_1)$. 
In addition, we obtain $m_2  = a \frac{n_1 n_2}{\mathrm{gcd}(n_2, d_1)} + m_1$, 
$m_3 =a \frac{n_2(n_1 + d_1)}{\mathrm{gcd}(n_2, d_1)} + m_1$. 
Thus we find that $C=a\frac{h \alpha_1n_1n_2}{(\alpha_2-\alpha_1) \mathrm{gcd}(d_1, n_2)} - h m_1$. 
The condition $0< C < h$ is expressed as 
$0<a\frac{\alpha_1n_1n_2}{(\alpha_2-\alpha_1) \mathrm{gcd}(d_1, n_2)} - m_1 < 1$, 
which allows many pairs of $(a, m_1)$. 
On the other hand, 
$t_p= \frac{a h n_1 n_2}{(\alpha_2 - \alpha_1)\mathrm{gcd}(d_1, n_2)}$ is proportional to $a$. 
In the $B=0$ case, we obtain 
many $t_p$. 
In the very small $|B| \neq 0$ case, because of the factor $e^{\frac{2B}{\hbar} t_p}$, 
the smallest $a$ should be chosen, 
and $m_1$, $m_2$, and $m_3$ are also determined. 
Thus the smallest $t_p$ is selected.

Finally let us consider the general case where the order of the Hilbert space is $n$ $(n=4, 5, \dots)$. 
We express  
$\mathrm{Re} \lambda_i ~(i \in A)$ as 
$\left\{ \mathrm{Re} \lambda_i \right\} =  \left\{ \alpha_1, \alpha_2, \dots, \alpha_n \right\}$, 
where we suppose that $\alpha_1 < \alpha_2 < \cdots < \alpha_n$.\footnote{In the case where there exists a subset $\left\{ i \right\}$ such that $\alpha_i = \alpha_{i+1}$, 
we just choose the integers $m_i$ and $m_{i+1}$ such that $m_i = m_{i+1}$ and $\alpha_i t_p - h m_i=C$ 
in the later argument.} 
In this case the condition (\ref{cond_Relambdait_p=hbarthetac2}) is expressed as 
$\alpha_1 t_p = C  +h m_1$, $\alpha_2 t_p = C  +h m_2$, $\dots$, 
$\alpha_n t_p = C  +h m_n$, 
where $m_1$, $m_2$, $\dots$, $m_n$ $(m_1 < m_2 < \cdots < m_n)$ 
are integers that are to be chosen properly, 
and $C$ is a constant $(0 \le C < h)$. 
In order for $t_p$ to obey the above relations, 
there have to exist integers $m_1$, $m_2$, $\dots$, $m_n$ that obey 
$\alpha_1 t_p - h m_1 = \alpha_2 t_p - h m_2 = \cdots  = \alpha_n t_p - h m_n= C$  
$\Leftrightarrow$ 
$t_p=\frac{h (m_{i+1} - m_i)}{\alpha_{i+1} -  \alpha_i}$ $(i=1,2, \dots, n-1)$ 
and $\frac{\alpha_2}{\alpha_1}=\frac{C+h m_2}{C+h m_1}$, which leads to 
$C=\frac{h}{\alpha_2 - \alpha_1}(\alpha_1 m_2 - \alpha_2 m_1)$. 
Let us suppose that the ratios 
of $\alpha_i -  \alpha_{i-1}$ to $\alpha_{i+1} -  \alpha_i$ $(i=2, 3, \dots, n-1)$ 
and of $\alpha_n -  \alpha_{n-1}$ to $\alpha_{n} -  \alpha_1$ are 
rational numbers\footnote{In the case where both of them are not rational numbers, 
we approximate the irrational numbers 
to rational ones in their neighborhoods, as we did in the previous case. }, and  
express them as 
$\frac{\alpha_i -\alpha_{i-1}}{\alpha_{i+1} -\alpha_i} =\frac{m_i -m_{i-1}}{m_{i+1} - m_i}=\frac{n_{i-1}}{d_{i-1}}$ 
$(i=2, 3, \dots, n-1)$  
and 
$\frac{\alpha_n -\alpha_{n-1}}{\alpha_n-\alpha_1} =\frac{m_n - m_{n-1}}{m_n - m_1}= \frac{n_{n-1}}{d_{n-1}}$, 
where $n_j$ and $d_j$ $(j=1,2, \dots, n-1)$ are positive and co-prime integers. 
Since we have the relations $(m_{i+2} -m_{i+1}) n_i= (m_{i+1} - m_i) d_i$ $(i=1, 2, \dots , n-2)$ 
and $(m_n-m_1)n_{n-1} = (m_n-m_{n-1})d_{n-1}$, we find 
$(m_{i+2} -m_{i+1}, m_{i+1} -m_i)= k_i (d_i, n_i)$  $(i=1, \dots , n-2)$, 
where $k_j$ $(j=1,2, \dots, n-1)$ are positive integers to be chosen properly. 
We are led to the relations $m_{i+1}-m_{i}=k_{i-1} d_{i-1} =k_{i} n_{i}$ $(i=2, \dots , n-1)$, 
so we find that the pairs $(k_i, k_{i+1})$ $(i=1,2, \dots, n-2)$ are expressed as 
$(k_i, k_{i+1})= \left\{ a_i /\mathrm{gcd}(n_{i+1}, d_i) \right\}  (n_{i+1}, d_i )$ $(i=1, \dots , n-2)$, 
where $a_j$ $(j=1,2, \dots, n-2)$ are positive integers to be chosen properly. 
Then we obtain $k_1=a_1 \frac{n_2}{\mathrm{gcd}(n_2, d_1)}$, 
$k_i=a_{i-1} \frac{d_{i-1}}{\mathrm{gcd}(n_i, d_{i-1})}=a_i \frac{n_{i+1}}{\mathrm{gcd}(n_{i+1}, d_i)}$ 
$(i=2, 3, \dots, n-2)$, 
and $k_{n-1}=a_{n-2} \frac{d_{n-2}}{\mathrm{gcd}(n_{n-1}, d_{n-2})}$. 
These representations suggest that we have to choose 
$a_1= l n_3 \frac{\mathrm{gcd}(n_2, d_1)}{\mathrm{gcd}(n_3, d_1)}$, 
$a_2= l d_1 \frac{\mathrm{gcd}(n_3, d_2)}{\mathrm{gcd}(n_3, d_1)}$, and 
$a_i= l \frac{\Pi_{k=1}^{i-1} d_k \mathrm{gcd}(n_{i+1}, d_i)}{\Pi_{l=4}^{i+1} n_l \mathrm{gcd}(n_3, d_1)}$ 
$(i=3, 4, \dots, n-2)$, 
which lead to  
$k_1=l\frac{n_2 n_3}{\mathrm{gcd}(d_1,n_3)}$, $k_2=l \frac{d_1 n_3}{\mathrm{gcd}(d_1,n_3)}$, 
$k_3=l\frac{d_1 d_2}{\mathrm{gcd}(d_1,n_3)}$, and 
$k_i= l \frac{\Pi_{k=1}^{i-1} d_k}{\Pi_{l=4}^{i} n_l \mathrm{gcd}(n_3, d_1)}$ 
$(i=4, 5, \dots, n-1)$, where $l$ is a positive integer to be chosen properly. 
Then, since $m_2=m_1+l\frac{n_1 n_2 n_3}{\mathrm{gcd}(n_3, d_1)}$, 
we find $C=h\left[ \frac{l n_1 n_2 n_3 \alpha_1}{(\alpha_2 - \alpha_1) \mathrm{gcd}(d_1,n_3) } - m_1 \right]$, 
and the condition $0<C<h$ is expressed as 
$0< \frac{l n_1 n_2 n_3 \alpha_1}{(\alpha_2 - \alpha_1) \mathrm{gcd}(d_1,n_3) } - m_1 < 1$, 
which allows many pairs of $(l, m_1)$. 
On the other hand, 
we find $t_p= hl \frac{n_1 n_2 n_3}{(\alpha_2 - \alpha_1) \mathrm{gcd}(n_3, d_1)}$, 
which is proportional to $l$. 
In the $B=0$ case, we obtain many $t_p$.  
In the very small $|B| \neq 0$ case, because of the factor $e^{\frac{2B}{\hbar} t_p}$, 
the smallest $l$ obeying the above inequality should be chosen, 
and $m_1$ and $m_i=\sum_{j=1}^{i-1} k_j n_j + m_1$ $(i =2, \dots, n)$ 
are also determined. Thus the smallest $t_p$ is selected.\footnote{The larger $n$ is, the larger the selected $t_p$ becomes.}  
Furthermore, in the case where the order of the Hilbert space is infinite,  
we can imagine obtaining selected $t_p$ similarly 
by considering the infinite limit of $n$ in the case where the order is $n$.

Now that we have proven the existence of $t_p$ such that the condition (\ref{cond_Relambdait_p=hbarthetac2}) is satisfied and so 
$\lvert \mathrm{Tr}  \left(   e^{-\frac{i}{\hbar} \hat{H} t_p}  \right) \rvert$ is maximized, 
$\langle \hat{\mathcal O} \rangle_{\mathrm{periodic}~\mathrm{time}}$ defined 
by Eq.(\ref{Ohat_periodic}) has been found for such $t_p$ to be reduced to the 
simpler expression given on the right-hand side of Eq.(\ref{Ohat_periodic_new}), which is real for 
$Q$-Hermitian $\hat{\mathcal O}$. 
Thus we have proven Theorem 3. 
Without considering the maximization principle, 
we do not have reality\footnote{ 
In the special case where the order of the Hilbert space 
labeled by the subset $A$ is just one, Theorem 2 is applied and Eq.(\ref{Ohat_periodic_new}) 
corresponds to Eq.(\ref{Ohat_periodic_new_order_of_A_is_1}). } for 
$\langle \hat{\mathcal O} \rangle_{\mathrm{periodic}~\mathrm{time}}$. 
In Theorem 3 there can be many states that are degenerate 
with regard to the imaginary parts of the eigenvalues of $\hat{H}$, 
so Theorem 3 is highly nontrivial even in the $B=0$ case 
compared to Theorem 2 by including the RAT. 
Finally we provide a corollary as the simplest case in Theorem 3, where a given Hamiltonian 
is Hermitian. 


\vspace*{0.5cm}

\noindent 
{\bf Corollary of Theorem 3} 
{\em Assume that a given Hamiltonian $\hat{H}$ is Hermitian. 
Then, provided that an operator $\hat{\mathcal O}$ is Hermitian, 
$\hat{\mathcal O}^{\dag} = \hat{\mathcal O}$, 
$\langle \hat{\mathcal O} \rangle_{\mathrm{periodic}~\mathrm{time}} \equiv \frac{\mathrm{Tr} \left(   e^{-\frac{i}{\hbar} \hat{H} t_p }  \hat{\mathcal O} \right) }{ \mathrm{Tr}  \left(   e^{-\frac{i}{\hbar} \hat{H} t_p}  \right) }$ 
becomes real for selected periods $t_p$'s 
such that $|\mathrm{Tr}  \left(   e^{-\frac{i}{\hbar} \hat{H} t_p}  \right)|$ is maximized. }


\vspace*{10mm}
%
\noindent 
{\bf\large  5. Discussion} \\
\noindent 
In this letter, after briefly reviewing our previous works, 
we studied the periodic complex action theory (CAT) that is obtained by imposing a periodic condition 
on the past and future states in the future-included CAT 
whose path runs over not only past but also future. 
In the periodic CAT, 
extending a normalized matrix element of an operator $\hat{\mathcal O}$, 
which is called the weak value in the RAT, 
to an expression such that various normalized matrix elements of $\hat{\mathcal O}$ 
are summed up with the weight of transition amplitudes, 
we introduced in Eq.(\ref{Ohat_periodic}) another normalized quantity 
$\langle \hat{\mathcal O} \rangle_{\mathrm{periodic}~\mathrm{time}}$ that is generically complex 
but expected to have a role of 
an expectation value for $\hat{\mathcal O}$. 
Seeking a condition for $\langle \hat{\mathcal O} \rangle_{\mathrm{periodic}~\mathrm{time}}$ to be real, 
we presented two theorems that hold in special cases. 
For a given period $t_p$ that is supposed to be sufficiently large from a phenomenological point of view, 
eigenstates of the Hamiltonian $\hat{H}$ that belong to the subset $A$ 
contribute significantly in the traces in the expression of 
$\langle \hat{\mathcal O} \rangle_{\mathrm{periodic}~\mathrm{time}}$ in Eq.(\ref{Ohat_periodic}), 
and thus $\langle \hat{\mathcal O} \rangle_{\mathrm{periodic}~\mathrm{time}}$  
is reduced to a simpler expression of Eq.(\ref{Ohat_periodic_new_A}).

In the first theorem (Theorem 2), considering a special case where the order of 
the subset $A$ is just one, i.e., the number of eigenstates that have the maximal imaginary part $B$ 
of the eigenvalues of $\hat{H}$ is just one, we claimed that 
for a given period $t_p$ 
the normalized quantity 
$\langle \hat{\mathcal O} \rangle_{\mathrm{periodic}~\mathrm{time}}$
becomes real,  
provided that $\hat{\mathcal O}$ is $Q$-Hermitian, i.e., 
Hermitian with regard to the modified inner product $I_Q$ 
that makes a given non-normal Hamiltonian $\hat{H}$ normal. 
In this case, both the numerator and denominator in the expression 
of Eq.(\ref{Ohat_periodic_new_A}) are dominated by 
the contribution from just a single eigenstate of $\hat{H}$, so 
phase factors in both cancel each other.  
Thus, in this special case, we obtained the expression of 
$\langle \hat{\mathcal O} \rangle_{\mathrm{periodic}~\mathrm{time}}$ 
in Eq.(\ref{Ohat_periodic_new_order_of_A_is_1}), and proved the theorem.

In the second theorem (Theorem 3), we considered another special case 
where $B \leq 0$ and $|B|$ is much smaller 
than the distances between any two real parts of the eigenvalues of  $\hat{H}$, 
and claimed that, 
provided that $\hat{\mathcal O}$ is $Q$-Hermitian,  
$\langle \hat{\mathcal O} \rangle_{\mathrm{periodic}~\mathrm{time}}$ 
becomes real for the period $t_p$ selected such that  
the absolute value of the transition amplitude 
$|\mathrm{Tr}  \left(   e^{-\frac{i}{\hbar} \hat{H} t_p}  \right)|$ is maximized. 
We proved via a number-theoretical argument 
that this theorem holds except for the special case where the order of the Hilbert space 
labeled by the subset $A$ is just one.\footnote{In this special case, 
$\langle \hat{\mathcal O} \rangle_{\mathrm{periodic}~\mathrm{time}}$ 
is found to be real for a given $t_p$, as is shown by Theorem 2.} 
In the other generic cases where the order of the Hilbert space is 
equal to or larger than two, 
we showed that such $t_p$ exist, for which 
$\langle \hat{\mathcal O} \rangle_{\mathrm{periodic}~\mathrm{time}}$ becomes real.
We argued that even the period $t_p$ 
can become an adjustment parameter to be determined 
via such a variant type of the maximization principle that we proposed 
in Refs.~\cite{Nagao:2015bya,Nagao:2017cpl}. 
This theorem suggests that, if our universe is periodic, then even 
the period could be fixed by our principle in the Feynman path integral.

In the study in this letter, we have supposed that the whole universe is a closed time-like curve (CTC) 
and considered very special cases for simplicity.   
However, in general relativity, more complicated universes can be considered. 
It would be interesting to investigate such more intricate ones. 
Now we might have a question: can we propose any model 
that would lead to a periodic universe in practice via any kind of maximization principle? 
It would be intriguing if we could propose a model that 
results in an exactly periodic universe 
and also provides for $t_p$ an order of magnitude identifiable with the age of our universe 
via any kind of maximization principle. 
In order to construct such a realistic model, it would be necessary to investigate 
the dynamics of the CAT in detail in some simple models.  
In Ref.~\cite{Nagao:2019dew} we formulated a harmonic oscillator model 
by introducing the two-basis formalism in the future-included CAT. 
It would be important to study the model further in detail.
Furthermore, since our $\langle \hat{\mathcal O} \rangle_{\mathrm{periodic}~\mathrm{time}}$ 
is independent of a reference time $t$, it would also be interesting to provide it 
with the time $t$ dependence by introducing a clock variable $T_\mathrm{clock}(t)$ 
to be inserted in the quantity. 
For this purpose we need to extend our series of reality theorems so that they hold 
not only for a single operator $\hat{\mathcal O}$ but also for a product of operators. 
We would like to report such investigations 
in the future (K.~Nagao and H.~B.~Nielsen, work in progress).



\ack

This work was supported by JSPS KAKENHI Grant Number JP21K03381, and accomplished 
partly during K.N.'s sabbatical stay in Copenhagen. 
He would like to thank the members and visitors of NBI 
for their kind hospitality and Klara Pavicic for her various kind arrangements and 
consideration during his visits to Copenhagen.  
H.B.N. is grateful to NBI for allowing him to work there as emeritus.

\let\doi\relax


\end{document}